\documentclass[pre, showpacs, preprint]{revtex4}
\usepackage{paralist}
\usepackage{natbib}
\usepackage{amsmath}
\usepackage{amsthm}
\usepackage{amssymb}
\usepackage{psfrag}
\usepackage{graphicx}
\usepackage{subfigure}

\newcommand{\NN}{\mathbb{N}}

\newcommand{\PP}{\mathbb{P}}

\newcommand{\Att}{A}
\newcommand{\probi}[2]{\PP_{#1}(#2\mid \tau_e)}
\newcommand{\prob}[2]{\PP_{#1}(#2)}
\newcommand{\ccdf}[2]{F^{c}_{#1}(#2)}
\newcommand{\transprob}[1]{W_{#1}}
\newcommand{\transprobq}[1]{W'_{#1}}

\newcommand{\Ordo}[1]{O\left(#1\right)}

\newcommand{\mean}[2]{\mathbb{E}_{#1}\left\{#2\right\}}
\newcommand{\meanq}[2]{\mathbb{E}_{#1}\left\{#2\mid q\right\}}
\newcommand{\meann}[2]{\mathbb{E}_{#1}\left\{#2\mid n\right\}}
\newcommand{\poch}[2]{\left(#1\right)_{#2}}
\newcommand{\sumterm}[1]{\Phi_\alpha(#1)}

\begin{document}
\title{Distribution of Edge Load in Scale-free Trees}
\author{Attila Fekete}
\email{fekete@complex.elte.hu}
\author{G\'abor Vattay}
\email{vattay@complex.elte.hu}
\affiliation{Department of Physics of Complex Systems, E\"otv\"os University, 
  P\'azm\'any P. s\'et\'any 1/A., H-1117 Budapest, Hungary}
\author{Ljupco Kocarev}
\email{lkocarev@ucsd.edu}
\affiliation{Institute for Nonlinear Sciences, University of California, 
  9500 Gilman Drive, La Jolla, CA 92093, San Diego, USA}
\begin{abstract}
  Node betweenness has been studied recently by a number of authors, but until
  now less attention has been paid to edge betweenness.  In this paper, we
  present an exact analytic study of edge betweenness in evolving scale-free
  and non-scale-free trees. We aim at the probability distribution of edge
  betweenness under the condition that a local property, the in-degree of
  the ``younger'' node of a randomly selected edge, is known. En route to the
  conditional distribution of edge betweenness the exact joint distribution of
  cluster size and in-degree, and its one dimensional marginal distributions
  have been presented in the paper as well. From the derived probability
  distributions the expectation values of different quantities have been
  calculated. Our results provide an exact solution not only for infinite,
  but for finite networks as well. 
\end{abstract}
\maketitle

\section{Introduction}
In recent years, the statistical properties of complex networks have been 
extensively investigated by the physics community \cite{Strogatz01,
  AlbertBarabasi02, DorogovtsevMendes02,Newman03}. With the increasing
computing power of modern computers, analysis of large-scale networks
and databases has become possible. It has been shown that the degree
statistics of many natural and artificial networks follow power law.
Examples for such networks vary from social interconnections and
scientific collaborations \cite{Newman01} to the world-wide web
\cite{BarabasiAlbertJeong00} and the Internet
\cite{FaloutsosFaloutsosFaloutsos99, Pastor01}. These networks are
usually referred to as \emph{scale-free} networks, since the power law 
distribution indicates that there is no characteristic scale in these systems.

{ In the early 1960s Erd\H{o}s and R\'enyi (ER) introduced random graphs
that served as the first  mathematical model of complex networks
\cite{ErdosRenyi60}.}  In their model, the number of nodes is fixed and
connections are established randomly, with probability $p_{ER}$.  Although the
ER model leads to rich theory, it fails to predict the power law distributions
observed in scale-free networks. Barab\'asi and Albert (BA) proposed a more
suitable evolving model of these networks
\cite{BarabasiAlbert99,BarabasiAlbertJeong99}.  The BA model is also based on
the random graph theory, but involves two key principles in addition:
\begin{inparaenum}[(a)]
\item \emph{growth}, that is, the size of the network is increasing
  during development, and 
\item \emph{preferential attachment},
  that is, new network elements are connected to higher degree nodes with
  higher probability. 
\end{inparaenum}
In the BA model every new node connects to the core network with a fixed 
number of links $m$.

The study of complex networks usually deals with the structural properties of
networks, like degree distribution \cite{Bollobas81}, shortest path 
distribution \cite{SzaboAlavaKertesz02}, degree--degree correlations, 
clustering \cite{WattsStrogatz98}, etc. For
complex networks which involve a transport mechanism \emph{betweenness} is
the matter of importance. Roughly speaking, betweenness is the number
of shortest paths passing through a certain network element. For example, in
communication networks information flows between remote hosts via
intermediate stations and in the Internet data packets are transmitted through 
routers and cables. The expected traffic flowing through a link or a
router is proportional to the particular edge or node betweenness, respectively. 
News and rumors spread in social networks, and node betweenness
measures the importance or centrality of an individual in society.

Node betweenness has been studied recently by \citeauthor*{GohKahngKim01}
\cite{GohKahngKim01}, who argued that it follows power law in scale-free
networks, and the exponent $\delta\approx2.2$ is independent from the degree
distribution in a certain range.  \citeauthor*{SzaboAlavaKertesz02}
\cite{SzaboAlavaKertesz02} used rooted deterministic trees to model scale-free
trees, and have found scaling exponent $\delta_t=2$. The same scaling exponent
has been found experimentally by \citeauthor*{GohKahngKim01} for scale-free
trees.  The rigorous proof of the heuristic results of
\cite{SzaboAlavaKertesz02} has been provided by \citeauthor*{BollobasRiordan04}
in \cite{BollobasRiordan04}.

Until recently, less attention has been paid to edge betweenness, even though
edge betweenness is often essential for estimating the load on links in complex
networks. For example, the edge betweenness can measure the ``importance'' of
relationships in social networks, or it can measure the expected amount of data
flow on links in computer networks. The probability distribution of edge
betweenness gives a rough statistical description of links and it characterizes
the network as a whole. Therefore, it is an important tool for an overall
description of links in complex networks.  

In some cases, a local property of the network is known as well. For instance,
if the number of friends of any individual can be counted, then it is
reasonable to ask the ``importance'' of a relationship (i.e. an edge in a
social network) under the condition that the number of friends of the related
individuals is known.  In this case, the \emph{conditional} probability
distribution of edge betweenness provides a much finer description of links
than the total distribution. 

In this paper we focus on how additional local information could be used to
describe links. In particular, we aim at deriving the probability distribution
of edge betweenness in evolving scale-free trees, under the condition that the
in-degree of the ``younger'' node of any randomly selected link is known.  For
the sake of simplicity we consider the in-degree of the ``younger'' node only.
Whether a node is ``younger'' than another node or not can be defined uniquely
in evolving networks, since nodes attach to the network sequentially. Note
that the in-degree is considered instead of total degree for practical reasons
only.  The construction of the network implies that the in-degree is less than
the total degree by one for every ``younger'' node.

To obtain the desired conditional distribution we calculate the exact joint
distribution of cluster size and in-degree for a \emph{specific} link first.
Then, the joint distribution of a \emph{randomly selected} link is derived,
which is comparable with the edge ensemble statistics obtained from a network
realization. The exact marginal distributions of cluster size and in-degree
follow next. After that, we give the distribution and mean of cluster size
under the condition that in-degree is known. For the sake of completeness the
conditional in-degree distribution is presented as well. Finally, the
distribution and mean of edge betweenness is derived under the condition that
the corresponding in-degree is known. Note that all of our analytic results are
\emph{exact even for finite networks}, which is valuable since the size of the
real networks are often much smaller than the valid range of asymptotic
formulae. Moreover, \emph{exact results for unbounded networks} are provided as
well. 

As a model of evolving scale-free trees we consider the BA model with parameter
$m=1$, extended with initial attractiveness \cite{Szymanski87,
DorogovtsevMendesSamukhin00}. With the initial attractiveness the scaling
properties of the network can be finely tuned. Note, that in the limit of
initial attractiveness to infinity the preferential attachment disappears, and
new nodes are connected to the old ones with uniform probability.  In this
limit the network looses its scale-free nature and becomes similar to an ER
network with $p_{ER}=2/N$. Therefore, scale-free and non-scale free networks
can be compared within one model. For the sake of simplicity the infinite limit
of { initial attractiveness} is referred to as the ``ER limit'' throughout
this paper.

We restrict our model to trees, that is to connected loopless graphs. The simplicity
of trees allows analytic results for edge betweenness, since the shortest paths in trees 
are unique between any pair of nodes. Although trees are special graphs, 
a number of real networks can be modelled by trees or by tree-like graphs with only a
negligible number of shortcuts. Important examples of such networks are the 
Autonomous Systems in the Internet \cite{CaldarelliMarchettiPietronero00}. 

The rest of this paper is organized as follows: In Section~\ref{sec:model}, a
short introduction to the construction of BA trees is given. Then, a master
equation for the joint distribution of cluster size and in-degree of a specific
edge is derived and solved in Section~\ref{sec:cond_joint_prob} and
Section~\ref{sec:sol_master}, respectively. The total joint distribution of
cluster size is calculated in Section~\ref{sec:joint_prob}. The marginal
and conditional distributions of cluster size and in-degree are derived in
Section~\ref{sec:marginal_prob} and Section~\ref{sec:cond_prob}, respectively.
In Section~\ref{sec:load_prob}, the conditional distribution of edge
betweenness follows. Finally, we conclude our work and discuss future
directions in Section~\ref{sec:conclusions}.

\section{The network model}
\label{sec:model}

The concepts of graph theory are used throughout this paper. 
A graph consists of \emph{vertices} (nodes) and \emph{edges} (links). Edges are
ordered or un-ordered pairs of vertices, depending on whether an ordered or
un-ordered graph is considered, respectively. The \emph{order} of a graph is
the number of vertices it holds, while the \emph{degree} of a vertex counts the
number of edges adjacent to it. \emph{Path} is also defined in the most natural
way: it is a vertex sequence, where any two consecutive elements form an edge.
A path is called a \emph{simple path} if none of the vertices in the path are
repeated. Any two vertices in a \emph{tree} can be connected by a unique simple
path.  The graph is called connected if for any vertex pair there exists a
path which starts from one vertex and ends at the other.

The construction of the network proceeds in discrete time steps. Let us denote
time with $\tau\in\NN$, and the developed graph with
$G_{\tau}=\left(V_{\tau},E_{\tau}\right)$, where $V_{\tau}$ and $E_{\tau}$
denote the set of vertices and the set of edges at time step $\tau$,
respectively.  Initially, at $\tau=0$, the graph consists only of a single
vertex without any edges.  Then, in every time step, a new vertex is connected
to the network with a single edge. The edge is \emph{directed}, which emphasize
that the two sides of the edge are not symmetric. The newly connected node,
which is the source of the edge, is always ``younger'' than the target node.
The term ``younger node of a link'' is used in this sense below. Note that the
initial vertex is different from all the others, since it has only incoming
connections; we refer to it as the \emph{root vertex}. 

The target of every new edge
is selected randomly from the present vertices of the graph. The probability
that a new vertex connects to an old one is proportional to the attractiveness
of the old vertex $v$, defined as
\begin{equation}
  \Att(v)=a+q,
\end{equation}
where parameter $a>0$ denotes the initial attractiveness and $q$ is the
in-degree of vertex $v$. { It has been shown in
\cite{DorogovtsevMendesSamukhin00} that the in-degree distribution is
asymptotically
$\PP(q)\simeq\left(1+a\right)\frac{\Gamma(2a+1)}{\Gamma(a)}\left(q+a\right)^{-\left(2+a\right)}$.
We will improve this result and derive the exact in-degree distribution below.}
Note that in the special case $a=0$ the attractiveness of every node is zero
except of the root vertex. It follows that every new vertex is connected to the
initial vertex in this case, which corresponds to a star topology. The special
case $a=1$ practically returns the original BA model.  Indeed, except for the
root vertex, the attractiveness of every vertex becomes equal to its degree if
$a=1$; this is exactly the definition of the attractiveness in the BA model
\cite{BarabasiAlbert99}. Finally, if $a\to\infty$, then preferential attachment
disappears in the limit, and the model tends to a Poisson-type graph, similar
to an ER graph.

The attractiveness of sub-graph $S$ is the sum of the attractiveness
of its elements:
\begin{equation}
  \Att(S)=\sum_{v'\in S}\Att(v').
\end{equation}
We refer to a connected sub-graph as a \emph{cluster}. The
attractiveness of cluster $C$ can be given easily:
\begin{equation}
  \Att(C)=
  \left(1+a\right)|C|-1,
\end{equation}
where $|C|$ denotes the size of the cluster. It is obvious that the
overall attractiveness of the network at time step $\tau$ is
\begin{equation}
  \Att(V_{\tau})=\left(1+a\right)\left(\tau+1\right)-1.
\end{equation}

\section{Master equation for the joint distribution of cluster size and in-degree}
\label{sec:cond_joint_prob}

\begin{figure}
  \begin{center}
    \psfrag{C}[c][c]{$C$}
    \psfrag{e}[c][c]{$e$}
    \psfrag{v}[c][c]{$v$}
    \psfrag{Root}[c][c]{Root}
    \resizebox{0.4\textwidth}{!}{\includegraphics[width=0.45\textwidth]{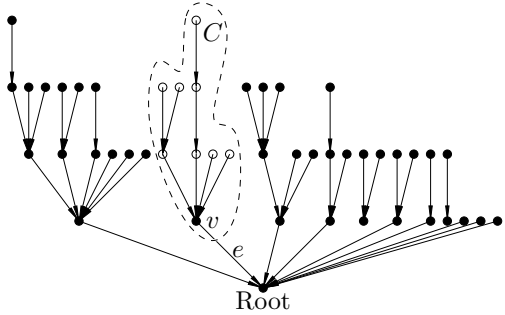}}
    \caption{(Color online) Schematic illustration of the evolving network at time $\tau$. 
      Vertex $v$, connected to the network at $\tau_e$, denotes the root
      of cluster $C$. Variables $q$ and $n=|C|-1$ denote the in-degree of vertex $v$ and 
      the number of nodes in $C$ without $v$ (marked by circles), respectively.}
    \label{fig:BAmodel}
  \end{center}
\end{figure}

Let us consider the size of the network $N$, an arbitrary edge $e$, which
connected vertex $v$ to the graph at time step $\tau_e>0$, and let us denote by
$C$ the cluster which has developed on vertex $v$ until $\tau>\tau_e$
(Fig.~\ref{fig:BAmodel}).  The calculation of betweenness of the given
edge is straightforward in trees, since the number of shortest paths going
through the given edge, that is the betweenness of the edge, is obviously
$L=|C|\left(N-|C|\right)$. Therefore, it is sufficient to know the size of the
cluster on the particular edge to get edge betweenness.  

The development of cluster $C$ can be regarded as a Markov process. The states
of the cluster are indexed by $\left(n,q\right)$, where $n=|C|-1$ denotes the
number of vertices in cluster $C$ without $v$. The in-degree of vertex $v$ is
denoted by $q$. Transition probabilities can be obtained from the definition of
preferential attachment: 
\begin{align}
  \transprob{\tau,n,q}&=\frac{\Att\left(C_{\tau}\setminus v\right)}{\Att\left(V_{\tau}\right)}
  =\frac{n-\alpha q}{\tau+1-\alpha}\\
  \transprobq{\tau,q}&=\frac{\Att\left(v\right)}{\Att\left(V_{\tau}\right)}
  =\frac{\alpha q+1-\alpha}{\tau+1-\alpha},
\end{align}
where $\alpha=1/\left(1+a\right)\in\left]0,1\right]$ and 
$\transprob{\tau,n,q}$ denotes the transition probability 
$\left(n,q\right)\to\left(n+1,q\right)$, and
$\transprobq{\tau,q}$ denotes the transition probability 
$\left(n,q\right)\to\left(n+1,q+1\right)$, respectively. 

The Master-equation, which describes the Markov process, follows from the fact 
that cluster $C$ can develop to state $\left(n,q\right)$ obviously 
in three ways: a new vertex can be connected 
\begin{enumerate}
\item to cluster $C$ but not to vertex $v$, and the cluster was in state $\left(n-1,q\right)$,
\item to vertex $v$, and the cluster was in state $\left(n-1,q-1\right)$, or
\item to the rest of the network, and the cluster was in state $\left(n,q\right)$.
\end{enumerate}
Therefore, the conditional probability $\probi{\tau}{n,q}$ that the developed 
cluster on edge $e$ is in state $\left(n, q\right)$ satisfies the following 
Master-equation: 
\begin{align}
  \label{eq:master_eq}
  \probi{\tau}{n,q}
  &=\transprob{\tau-1,n-1,q}\,\probi{\tau-1}{n-1,q}\\
  \notag
  &+\transprobq{\tau-1,q-1}\probi{\tau-1}{n-1,q-1}\\
  \notag
  &+\left[1-\transprob{\tau-1,n,q}-\transprobq{\tau-1,q}\right]\probi{\tau-1}{n,q},
\end{align}
Since the process starts with $n=0$, $q=0$ at $\tau=\tau_e$, 
the initial condition of the above Master equation is
$\probi{\tau_e}{n,q}=\delta_{n,0}\delta_{q,0}$, where $\delta_{i,j}$ is the
Kronecker-delta symbol. 

\section{The solution of the master equation}
\label{sec:sol_master}

After substituting the above transition probabilities into \eqref{eq:master_eq},
the following first order linear partial difference equation is obtained:
\begin{align}
  \notag
  \left(\tau-\alpha\right)\,\probi{\tau}{n,q}
  &=\left(n-1-\alpha q\right)\probi{\tau-1}{n-1,q}\\
  \notag
  &+\left(\alpha q+1-2\alpha\right)\probi{\tau-1}{n-1,q-1}\\
  \label{eq:master_eq_re}
  &+\left(\tau-n-1\right)\probi{\tau-1}{n,q},
\end{align} 

Let us seek a particular solution of \eqref{eq:master_eq_re}
in product form: $f(\tau)\,g(n)\,h(q)$. The following equation is obtained 
after substituting the probe function into \eqref{eq:master_eq_re}:
\begin{align*}
	\left(\tau-\alpha\right)\frac{f(\tau)}{f(\tau-1)}-\tau
	&=\left(n-1-\alpha q\right)\frac{g(n-1)}{g(n)}-n-1
	+\left(\alpha q+1-2\alpha\right)\frac{g(n-1)}{g(n)}\frac{h(q-1)}{h(q)}.
\end{align*}
The above partial difference equation can be separated into a system of three
ordinary difference equations. The solutions of the separated equations are:
\begin{align}
  f(\tau)&=\frac{\Gamma(\tau+\lambda_1)}{\Gamma(\tau-\alpha+1)},\\
  g(n)&=\frac{\Gamma(n+\lambda_2)}{\Gamma(n+\lambda_1+1)},\\
  h(q)&=\frac{\Gamma(q+1/\alpha-1)}{\Gamma(q+\lambda_2/\alpha+1)},
\end{align}
where $\lambda_1$ and $\lambda_2$ are separation parameters.

The solution of \eqref{eq:master_eq}, which fulfils the
initial conditions, is constructed from the linear combination
of the above particular solutions:
\begin{equation}
	\probi{\tau}{n,q}=\sum_{\lambda_1,\lambda_2}
	C_{\lambda_1,\lambda_2}\,f(\tau)\,g(n)\,h(q),
\end{equation}
where $C_{\lambda_1,\lambda_2}$ coefficients are independent of 
$\tau$, $n$ and $q$.

To obtain coefficients $C_{\lambda_1,\lambda_2}$, the initial condition of
\eqref{eq:master_eq} is expanded on the bases of $g(n)$ and $h(q)$. The
detailed calculation is presented in Appendix~\ref{app:kronecker_exp}. 

The solution of \eqref{eq:master_eq} is
	\begin{equation}
		\probi{\tau}{n,q}=\frac{\Gamma(\tau-\tau_e+1)}{\Gamma(\tau_e)\,\Gamma(n+1)}
		\frac{\Gamma(\tau-n)}{\Gamma(\tau-\tau_e-n+1)}
		\frac{\Gamma(\tau_e+1-\alpha)}{\Gamma(\tau+1-\alpha)}
		\frac{\Gamma(q+1/\alpha-1)}{\Gamma(1/\alpha-1)}\,
		\sumterm{n,q}
		\label{eq:master_eq_sol_i}
	\end{equation}
where $\sumterm{n,q}=\sum_{k=0}^{q}\frac{\left(-1\right)^k}{k!\left(q-k\right)!}\poch{-\alpha k}{n}$
and $\poch{x}{n}\equiv\Gamma(n+x)/\Gamma{(x)}$ denotes Pochhammer's symbol. Note that
$\probi{\tau}{n,q}\neq0$ iff $0\le q\le n\le\tau-\tau_e$. The conditions $0\le q$ and
$n\le\tau-\tau_e$ are obvious, since $1/\Gamma(k)=0$ by definition if $k$ is a negative
integer or zero. Furthermore, the condition $q<n$ can be easily seen if $\sumterm{n,q}$ 
is transformed into the following equivalent form: 
$\sumterm{n,q}=\frac{1}{q!}\frac{d^n}{dz^n}z^{n-1}\left(1-z^{-\alpha}\right)^q\bigr|_{z=1}$.
This result coincides with the fact that the size of a cluster $n$ cannot be less than
the corresponding number of in-degrees $q$.

\section{Joint distribution of cluster size and in-degree}
\label{sec:joint_prob}

Equation \eqref{eq:master_eq_sol_i} provides the conditional probability that a
particular edge which was connected to the network at $\tau_e$ is in state
$\left(n,q\right)$ at $\tau>\tau_e$.  In a fully developed network, however,
the time when a particular edge is connected to the network is usually not
known. Moreover, the development of an individual link is usually not as important 
as the properties of the finally developed link ensemble. Therefore, we are more
interested in the total probability $\prob{\tau}{n,q}$, that is the probability that
a randomly selected edge is in state $\left(n,q\right)$ at $\tau$, than the
conditional probability \eqref{eq:master_eq_sol_i}. The total probability can be 
calculated with the help of the total probability theorem:
\begin{equation}
	\prob{\tau}{n,q}=\sum_{\tau_e=1}^{\tau}\probi{\tau}{n,q}\,
	\prob{\tau}{\tau_e},
\end{equation}
where $\prob{\tau}{\tau_e}$ is the probability that a randomly selected 
edge was included into the network at $\tau_e$. According to the construction 
of the network one edge is added to the network at every time step, therefore
$\prob{\tau}{\tau_e}=1/\tau$. The following formula can be obtained after
the above summation has been carried out:
\begin{equation}
	\prob{\tau}{n,q}=
	\frac{\tau+1-\alpha}{\tau}\frac{\poch{1/\alpha-1}{q}}{\poch{2-\alpha}{n+1}}
	\,\sumterm{n,q},
	\label{eq:sol_joint}
\end{equation}
where $0<\alpha\le1$. In star topology, that is when $\alpha=1$,
the joint distribution $\prob{\tau}{n,q}$
evidently degenerates to $\prob{\tau}{n,q}=\delta_{n,0}\,\delta_{q,0}$.

The ER limit of joint distribution can be obtained via the $\alpha\to0$ limit of
\eqref{eq:sol_joint} (see Appendix~\ref{app:ER_limit} for details):
\begin{align}
  \lim_{\alpha\to0}\prob{\tau}{n,q}&=\frac{\tau+1}{\tau}
  \sum_{k=q-1}^{n-1}\left(-1\right)^{k+n-1}
  \frac{\binom{k}{q-1}S_{n-1}^{\left(k\right)}}{\Gamma(n+3)}
  \label{eq:sol_joint_a0}
\end{align}
where $0<q\le n<\tau$ and $S_n^{\left(m\right)}$ denote the Stirling numbers of
the first kind. Note, that for the special case $n=q=0$ the ER limit is
$\lim_{\alpha\to0}\prob{\tau}{0,0}=\frac{\tau+1}{2\tau}$.

\begin{figure}[tb]
  \begin{center}
    \psfrag{n}[c][c][1]{Cluster size, $n$}
    \psfrag{simulation}[l][l][1]{Simulation}
    \psfrag{analytic}[l][l][1]{Analytic}
    \psfrag{error}[c][c][1]{Relative difference}
    \psfrag{q}[c][c][1]{In-degree, $q$}
    \psfrag{q=1}[r][r][1]{$q=1$}
    \psfrag{q=10}[r][r][1]{$q=10$}
    \psfrag{q=20}[r][r][1]{$q=20$}
    \psfrag{n=5}[r][r][1]{$n=5$}
    \psfrag{n=10}[r][r][1]{$n=10$}
    \psfrag{n=20}[r][r][1]{$n=20$}
    \psfrag{n=40}[r][r][1]{$n=40$}
    \psfrag{P(n,q)}[c][c][1]{$\prob{\tau}{n,q}$}
    \psfrag{ 5}[c][c][1]{$5$}
    \psfrag{ 15}[c][c][1]{$15$}
    \psfrag{ 20}[c][c][1]{$20$}
    \psfrag{ 25}[c][c][1]{$25$}
    \psfrag{ 30}[c][c][1]{$30$}
    \psfrag{ 1}[l][l][1]{$1$}
    \psfrag{ 10}[l][l][1]{$10$}
    \psfrag{ 100}[l][l][1]{$100$}
    \psfrag{ 1000}[l][l][1]{$1000$}
    \psfrag{ 0.1}[l][l][1]{$10^{-1}$}
    \psfrag{ 0.01}[l][l][1]{$10^{-2}$}
    \psfrag{ 0.001}[l][l][1]{$10^{-3}$}
    \psfrag{ 1e-04}[l][l][1]{$10^{-4}$}
    \psfrag{ 1e-05}[l][l][1]{$10^{-5}$}
    \psfrag{ 1e-06}[l][l][1]{$10^{-6}$}
    \psfrag{ 1e-07}[l][l][1]{$10^{-7}$}
    \psfrag{ 1e-08}[l][l][1]{$10^{-8}$}
    \psfrag{ 1e-10}[l][l][1]{$10^{-10}$}
    \psfrag{ 1e-12}[l][l][1]{$10^{-12}$}
    \psfrag{ 1e-14}[l][l][1]{$10^{-14}$}
    \subfigure[Joint distribution of cluster size and in-degree as the function of cluster size.]{
    	\label{subfig:deg-cl-prob-n}
    	\resizebox{0.5\textwidth}{!}{\includegraphics{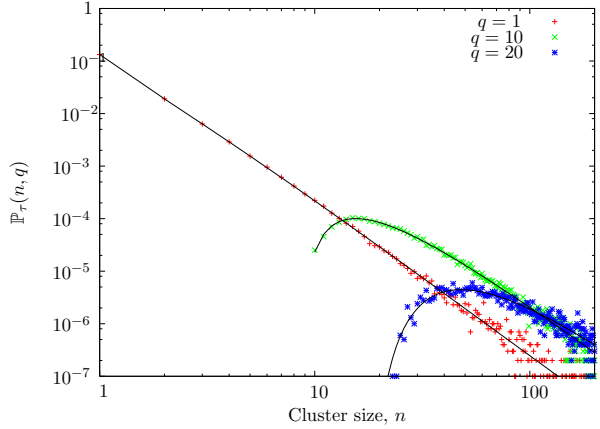}}
    }
    \subfigure[Joint distribution of cluster size and in-degree as the function of in-degree.]{
    	\label{subfig:deg-cl-prob-q}
    	\resizebox{0.5\textwidth}{!}{\includegraphics{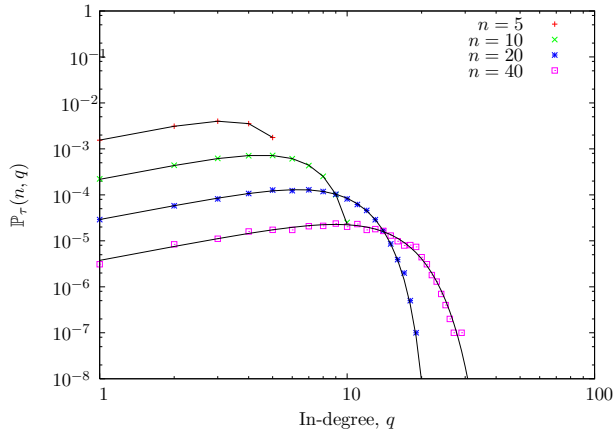}}
    }
  \end{center}
  \caption{(Color online)  Joint empirical distribution of cluster size and in-degree
  at $\alpha=1/2$ (symbols), and analytic formula
  (\ref{eq:sol_joint}) (solid lines) are compared 
  on double-logarithmic plot. Simulation results have been
  obtained from $100$ realizations of $10^5$ size networks. 
  }
  \label{fig:deg-cl-prob}
\end{figure}

The above formulae have been verified by extensive numerical simulations.  The
joint empirical cluster size and in-degree distribution has been compared with
the analytic formula \eqref{eq:sol_joint} for $\alpha=1/2$ in
Fig~\ref{fig:deg-cl-prob}.  Subfigures~\ref{subfig:deg-cl-prob-n}
and~\ref{subfig:deg-cl-prob-q} represent intersections of the joint
distribution with cutting planes of fixed in-degrees and cluster sizes,
respectively.  The figures confirm that the empirical distributions, obtained
as relative frequencies of links with cluster size $n$ and in-degree $q$ in 100
network realizations, are in complete agreement with the derived analytic
results. 

Equation~\eqref{eq:sol_joint} is the fundamental result of this section.  The
derived distribution is exact for any finite value of $\tau$, that is for any
finite BA trees. This result is precious for modeling a number of real
networks, where the size of the network is small, compared to the relevant
range of cluster size or in-degree. If the size of the network is much larger
than the relevant range of cluster size or in-degree, then it is practical to
consider the network as infinitely large, that is to take the $\tau\to\infty$
limit. For the above joint distributions \eqref{eq:sol_joint} and \eqref{eq:sol_joint_a0}
the $\tau\to\infty$ limit is evident, since the $\tau$ dependent prefactors 
obviously tend to $1$ if the size of the networks grow beyond every limit.

\section{Marginal distributions of cluster size and in-degree}
\label{sec:marginal_prob}

We have derived the joint probability distribution of the cluster size and the
in-degree in the previous section.  In many cases it is sufficient to know the
probability distribution of only one random variable, since the information on
the other variable is either unavailable or not needed. It is also possible
that the one dimensional distribution is needed especially, for example, for
the calculation of a conditional distribution in Section~\ref{sec:cond_prob}. 

The one dimensional (marginal) distributions $\prob{\tau}{n}$ and 
$\prob{\tau}{q}$ can be obtained from joint distribution $\prob{\tau}{n,q}$ 
as follows:
\begin{align*}
  \prob{\tau}{n}&=\sum_{q=0}^{n}\prob{\tau}{n,q}, &
  \prob{\tau}{q}&=\sum_{n=q}^{\tau-1}\prob{\tau}{n,q}.
\end{align*}
After substituting \eqref{eq:sol_joint} into the above formulae
the following expressions are obtained:
\begin{equation}
  \label{eq:sol_marginal_n}
  \prob{\tau}{n}=\frac{\tau+1-\alpha}{\tau}
  \frac{1-\alpha}{\left(n+1-\alpha\right)\left(n+2-\alpha\right)}.
\end{equation}
if $0\le n<\tau$ and $\prob{\tau}{n}=0$ if $n\ge\tau$. Furthermore,
\begin{align}
  \label{eq:sol_marginal_q}
  \prob{\tau}{q}=\frac{\tau+1-\alpha}{\tau}\frac{1}{\alpha}
  \frac{\poch{1/\alpha-1}{1/\alpha}}{\poch{q+1/\alpha-1}{1/\alpha+1}}
  -\frac{\tau+1-\alpha}{\tau}\frac{\poch{1/\alpha-1}{q}}{\poch{2-\alpha}{\tau}}
  \sum_{k=0}^{q}\frac{\left(-1\right)^k}{k!\left(q-k\right)!}\frac{\poch{-\alpha k}{\tau}}{\alpha k+2-\alpha}.
\end{align}
if $0\le q<\tau$ and $\prob{\tau}{q}=0$ otherwise. 
Rice's method \cite{Odlyzko95}
has been applied to evaluate the first term of $\prob{\tau}{q}$ in closed form.

The ER limit of the marginal cluster size distribution can 
obviously be obtained from \eqref{eq:sol_marginal_n} at $\alpha=0$.
Furthermore, the ER limit of the marginal in-degree distribution 
can be derived analogously to the limit of the joint distribution, 
shown in Appendix~\ref{app:ER_limit}:
\begin{equation}
  \lim_{\alpha\to0}\prob{\tau}{q}
  =\frac{\tau+1}{\tau}\frac{1}{2^{q+1}}
  +\frac{\tau+1}{\tau}\frac{1}{\Gamma(\tau+2)\Gamma(q)}\frac{d^{q-1}}{d\alpha^{q-1}}
  \frac{\poch{1+\alpha}{\tau-1}}{2-\alpha}\Biggr|_{\alpha=0}.
\end{equation}

If the size of the network grows beyond every limit, that is if $\tau\to\infty$,
then the marginal distributions become much simpler:
\begin{align}
  \prob{\infty}{n}&=\frac{1-\alpha}{\left(n+1-\alpha\right)\left(n+2-\alpha\right)}\\
  \prob{\infty}{q}&=\frac{1}{\alpha}\frac{\poch{1/\alpha-1}{1/\alpha}}
  {\poch{q+1/\alpha-1}{1/\alpha+1}}\\
  \lim_{\alpha\to0}\prob{\infty}{q}&=2^{-q-1}.
\end{align}

The asymptotic behavior of the cluster size and in-degree distributions differ
significantly. The tail of the cluster size distribution follows power law with
exponent $2$ either in BA or ER network, independently of $\alpha$. However, we
learned that the tail of the in-degree distribution follows power law with
exponent $1/\alpha+1=2+a$ in BA networks, and it falls exponentially in ER
topology, which agree with the well known results of previous works 
\cite{ErdosRenyi60}.

\begin{figure}[tb]
  \begin{center}
    \psfrag{n}[c][c][1]{Cluster size, $n$}
    \psfrag{q}[c][c][1]{In-degree, $q$}
    \psfrag{simulation}[l][l][1]{Simulation}
    \psfrag{analytic}[l][l][1]{Analytic}
    \psfrag{Fc(n)}[c][c][1]{$\ccdf{\tau}{n}$}
    \psfrag{Fc(q)}[c][c][1]{$\ccdf{\tau}{q}$}
    \psfrag{alpha=0}[r][r][1]{$\alpha=0$}
    \psfrag{alpha=1/3}[r][r][1]{$\alpha=1/3$}
    \psfrag{alpha=1/2}[r][r][1]{$\alpha=1/2$}
    \psfrag{alpha=2/3}[r][r][1]{$\alpha=2/3$}
    \psfrag{ 0}[r][r][1]{$0$}
    \psfrag{ 5}[r][r][1]{$5$}
    \psfrag{ 10}[r][r][1]{$10$}
    \psfrag{ 15}[r][r][1]{$15$}
    \psfrag{ 20}[r][r][1]{$20$}
    \psfrag{ 25}[r][r][1]{$25$}
    \psfrag{ 1}[r][r][1]{$1$}
    \psfrag{ 10}[r][r][1]{$10$}
    \psfrag{ 100}[r][r][1]{$10^2$}
    \psfrag{ 1000}[r][r][1]{$10^3$}
    \psfrag{ 10000}[r][r][1]{$10^4$}
    \psfrag{ 100000}[r][r][1]{$10^5$}
    \psfrag{ 1e+06}[r][r][1]{$10^6$}
    \psfrag{ 0.1}[r][r][1]{$10^{-1}$}
    \psfrag{ 0.01}[r][r][1]{$10^{-2}$}
    \psfrag{ 0.001}[r][r][1]{$10^{-3}$}
    \psfrag{ 1e-04}[r][r][1]{$10^{-4}$}
    \psfrag{ 1e-05}[r][r][1]{$10^{-5}$}
    \psfrag{ 1e-06}[r][r][1]{$10^{-6}$}
    \psfrag{ 1e-07}[r][r][1]{$10^{-7}$}
    \psfrag{ 1e-08}[r][r][1]{$10^{-8}$}
    \psfrag{ 1e-09}[r][r][1]{$10^{-9}$}
    \psfrag{ 1e-10}[r][r][1]{$10^{-10}$}
    \resizebox{0.5\textwidth}{!}{\includegraphics{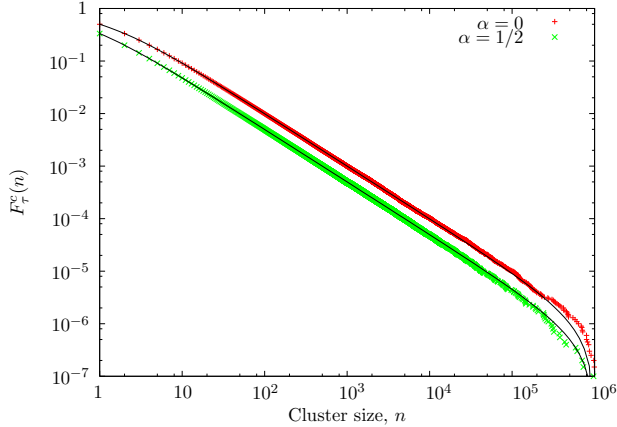}}
  \end{center}
  \caption{(Color online) Figure shows comparison of empirical CCDFs of cluster size  
  distributions (points) with analytic formula (\ref{eq:ccdf_n}) 
  (lines) on logarithmic plots, at $\alpha=0$, and $1/2$. Empirical 
  distributions have been obtained from $10$ realizations of $N=10^6$ size networks.}
  \label{fig:cluster-dist}
\end{figure}

It is worth noting that the mean cluster size diverges logarithmically as
the size of the network tends to infinity:
$\mean{\tau}{n}=\sum_{n=0}^{\tau-1}n\,\prob{\tau}{n}=\left(1-\alpha\right)\ln\tau+\Ordo{1}$.
The expectation value of the in-degree, however, obviously remains finite:
$\mean{\tau}{q}=\frac{\tau}{\tau+1}<1$,
and $\mean{\infty}{q}=1$ if the size of the network is infinite. Moreover, the
standard error of the in-degree can be also given exactly when the size of 
the network grows beyond every limit:
\begin{equation}
  \mean{\infty}{\left(q-1\right)^2}=\frac{2}{\left|1-2\alpha\right|}.
\end{equation}
This result implies that the fluctuations of the in-degree diverge
in a boundless network, if $\alpha=1/2$, that is in the classical BA model.

\begin{figure}[tb]
  \begin{center}
    \psfrag{n}[c][c][1]{Cluster size, $n$}
    \psfrag{q}[c][c][1]{In-degree, $q$}
    \psfrag{simulation}[l][l][1]{Simulation}
    \psfrag{analytic}[l][l][1]{Analytic}
    \psfrag{Fc(n)}[c][c][1]{$\ccdf{\tau}{n}$}
    \psfrag{Fc(q)}[c][c][1]{$\ccdf{\tau}{q}$}
    \psfrag{alpha=0}[r][r][1]{$\alpha=0$}
    \psfrag{alpha=1/3}[r][r][1]{$\alpha=1/3$}
    \psfrag{alpha=1/2}[r][r][1]{$\alpha=1/2$}
    \psfrag{alpha=2/3}[r][r][1]{$\alpha=2/3$}
    \psfrag{ 0}[r][r][1]{$0$}
    \psfrag{ 5}[r][r][1]{$5$}
    \psfrag{ 10}[r][r][1]{$10$}
    \psfrag{ 15}[r][r][1]{$15$}
    \psfrag{ 20}[r][r][1]{$20$}
    \psfrag{ 25}[r][r][1]{$25$}
    \psfrag{ 1}[r][r][1]{$1$}
    \psfrag{ 10}[r][r][1]{$10$}
    \psfrag{ 100}[r][r][1]{$10^2$}
    \psfrag{ 1000}[r][r][1]{$10^3$}
    \psfrag{ 10000}[r][r][1]{$10^4$}
    \psfrag{ 100000}[r][r][1]{$10^5$}
    \psfrag{ 0.1}[r][r][1]{$10^{-1}$}
    \psfrag{ 0.01}[r][r][1]{$10^{-2}$}
    \psfrag{ 0.001}[r][r][1]{$10^{-3}$}
    \psfrag{ 1e-04}[r][r][1]{$10^{-4}$}
    \psfrag{ 1e-05}[r][r][1]{$10^{-5}$}
    \psfrag{ 1e-06}[r][r][1]{$10^{-6}$}
    \psfrag{ 1e-07}[r][r][1]{$10^{-7}$}
    \psfrag{ 1e-08}[r][r][1]{$10^{-8}$}
    \psfrag{ 1e-09}[r][r][1]{$10^{-9}$}
    \psfrag{ 1e-10}[r][r][1]{$10^{-10}$}
    \resizebox{0.5\textwidth}{!}{\includegraphics{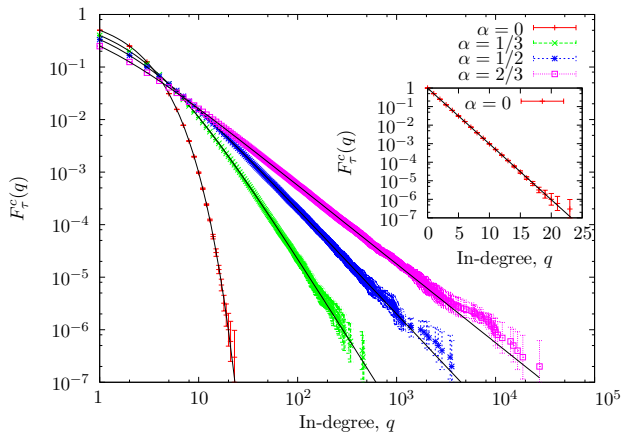}}
  \end{center}
  \caption{(Color online) Figure shows comparison of empirical CCDFs of in-degree 
  distributions (points) with analytic formula (\ref{eq:ccdf_q}) 
  (lines) on logarithmic plots, at $\alpha=0$, $1/3$, $1/2$, and $2/3$. Empirical 
  distributions have been obtained from $10$ realizations of $N=10^6$ size networks.
  Inset: Comparison at $\alpha=0$ on semi-logarithmic plot.}
  \label{fig:deg-dist}
\end{figure}

Our analytic results have been verified with computer simulations.  Since
cumulative distributions are more suitable to be compared with simulations than
ordinary distributions, we matched the corresponding complementary cumulative
distribution functions (CCDF) against simulation data. The CCDF of cluster
size, $\ccdf{\tau}{n}=\sum_{n'=n}^{\tau-1}\prob{\tau}{n'}$ can be calculated
straightforwardly:
\begin{equation}
  \ccdf{\tau}{n}
  =\frac{\tau+1-\alpha}{\tau}\frac{1-\alpha}{n+1-\alpha}
  -\frac{1-\alpha}{\tau},
  \label{eq:ccdf_n}
\end{equation}
where $0\le n< \tau$ and $0\le\alpha\le1$. 
The CCDF of in-degree, $\ccdf{\tau}{q}=\sum_{q'=q}^{\tau-1}\prob{\tau}{q'}$ 
is more complex, however:
\begin{align}
  \notag
  \ccdf{\tau}{q}
  &=\frac{\tau+1-\alpha}{\tau}\frac{\poch{1/\alpha-1}{1/\alpha}}{\poch{q+1/\alpha-1}{1/\alpha}}
  -\frac{1-\alpha}{\tau}\\
  &+\frac{\tau+1-\alpha}{\tau}\frac{\poch{1/\alpha-1}{q}}{\poch{2-\alpha}{\tau}}
  \sum_{k=0}^{q-2}\frac{\left(-1\right)^k}{k!\left(q-2-k\right)!}
  \frac{\poch{1-\alpha-\alpha k}{\tau-1}}{\left(k+1/\alpha\right)\left(k+2/\alpha\right)}
  \label{eq:ccdf_q}
 \end{align}
where $0\le q<\tau$ and $0<\alpha\le1$. If the size of the network grows
beyond every limit, then the CCDFs are the following:
\begin{align}
  \ccdf{\infty}{n}&=\frac{1-\alpha}{n+1-\alpha}, & 
  \ccdf{\infty}{q}&=
  \frac{\poch{1/\alpha-1}{1/\alpha}}{\poch{q+1/\alpha-1}{1/\alpha}},
\end{align}
where $0\le n$, $0\le q$ and $0<\alpha<1$.

Comparison of analytic CCDF of cluster size \eqref{eq:ccdf_n} and empirical
distributions are shown in Figure~\ref{fig:cluster-dist} for $\alpha=0$, $1/3$,
$1/2$, and $2/3$. Experimental data has been collected from $10$ realizations of
$10^6$ node networks.  Figure~\ref{fig:cluster-dist} shows that simulations
fully confirm our analytic result.

On Figure~\ref{fig:deg-dist} analytic formula \eqref{eq:ccdf_q} and the
empirical CCDFs of in-degree, obtained from the same $10^6$ node realizations,
are compared.  Note the precise match of the simulation and the theoretical
distribution on almost the whole data range. Some small discrepancy can be
observed around the low probability events.  This deviation is caused by the
aggregation of errors on the cumulative distribution when some rare event occurs
in a finite network.

\section{Conditional probabilities and expectation values}
\label{sec:cond_prob}

In the previous sections exact joint and marginal distributions of cluster size
and in-degree have been analyzed for both finite and infinite networks. All
these distributions provide general statistics of the network.  In this section
we proceed further, and we investigate the scenario when the ``younger''
in-degree of a randomly selected link is known. We ask the cluster size
distribution under this condition, that is the conditional distribution
$\prob{\tau}{n\mid q}$. The results of the previous sections are referred to
below to obtain the conditional probability distribution, and eventually the
conditional expectation of cluster size. For the sake of completeness, the
conditional distribution and expectation of in-degree are given as well at the
end of this section.

The conditional cluster size distribution can be given by
the quotient of the joint and the marginal in-degree distributions by definition:
\begin{equation}
  \prob{\tau}{n\mid q}=\frac{\prob{\tau}{n,q}}{\prob{\tau}{q}}.
\end{equation}
The exact conditional distribution for any finite network can be obtained after
substituting (\ref{eq:sol_joint}) and (\ref{eq:sol_marginal_q}) into the above
expression. For a boundless network the conditional distribution takes the simpler
form:
\begin{equation}
  \prob{\infty}{n\mid q}=\alpha\frac{\poch{2/\alpha-1}{q+1}}{\poch{2-\alpha}{n+1}}\sumterm{n,q},
  \label{eq:mean_cond_cl_inf}
\end{equation}
where $0\le q\le n$. If $n\gg1$, then 
$\prob{\infty}{n\mid q}\sim \alpha\poch{2/\alpha-1}{q+1}/n^{3}+\Ordo{1/n^4}$, that is
the conditional cluster size distribution falls faster than
the ordinary cluster size distribution. It follows that the mean
of the conditional cluster size distribution will not diverge like the
mean of the ordinary distribution.

What is the expected size of a cluster under the condition that the
in-degree of its root is known? For practical reasons, we do not 
calculate $\meanq{\tau}{n}$ directly, but we calculate
$\meanq{\tau}{n+2-\alpha}=\meanq{\tau}{n}+2-\alpha$ instead:
\begin{equation}
  \meanq{\tau}{n+2-\alpha}
  =\frac{1}{\prob{\tau}{q}}\sum_{n=q}^{\tau-1}\left(n+2-\alpha\right)\prob{\tau}{n,q}.
\end{equation}
Since $\left(n+2-\alpha\right)\prob{\tau}{n,q}=\frac{\tau+1-\alpha}{\tau}
\frac{\poch{1/\alpha-1}{q}}{\poch{2-\alpha}{n}}\sumterm{n,q}$, the above 
summation can be given similarly to the marginal distribution $\prob{\tau}{q}$ in 
(\ref{eq:sol_marginal_q}):
\begin{align*}
  \notag
  \sum_{n=q}^{\tau-1}\left(n+2-\alpha\right)\prob{\tau}{n,q}
  &=\frac{\tau+1-\alpha}{\tau}\frac{1/\alpha-1}{q+1/\alpha-1}\\
  &-\frac{\tau+1-\alpha}{\tau}\frac{\poch{1/\alpha-1}{q}}{\poch{2-\alpha}{\tau-1}}
  \sum_{k=0}^{q}\frac{\left(-1\right)^k}{k!\left(q-k\right)!}\frac{\poch{-\alpha k}{\tau}}{\alpha k+1-\alpha}
\end{align*}
After replacing the above sum in $\meanq{\tau}{n}$, the following equation can be obtained:
\begin{equation}
  \meanq{\tau}{n+2-\alpha}
  =\left(1-\alpha\right)\frac{\poch{q+1/\alpha}{1/\alpha}}{\poch{1/\alpha-1}{1/\alpha}}G_{\tau}(q),
  \label{eq:mean_q}
\end{equation}
where
\begin{equation}
  G_{\tau}(q)=\frac{\displaystyle 1-\frac{\poch{1/\alpha-1}{q+1}}{\poch{1-\alpha}{\tau}}
    \sum_{k=0}^{q}\frac{\left(-1\right)^k}{k!\left(q-k\right)!}\frac{\poch{-\alpha k}{\tau}}{k+1/\alpha-1}}
    {\displaystyle 1-\frac{\poch{2/\alpha-1}{q+1}}{\poch{2-\alpha}{\tau}}
    \sum_{k=0}^{q}\frac{\left(-1\right)^k}{k!\left(q-k\right)!}\frac{\poch{-\alpha k}{\tau}}{k+2/\alpha-1}}.
\end{equation}
The identity $\lim_{\tau\to\infty}G_{\tau}(q)\equiv1$ implies that
$G_{\tau}(q)$ involves the finite scale effects, and the factors preceding
$G_{\tau}(q)$ give the asymptotic form of $\meanq{\tau}{n+2-\alpha}$: 
\begin{equation}
  \meanq{\infty}{n+2-\alpha}
  =\left(1-\alpha\right)\frac{\poch{q+1/\alpha}{1/\alpha}}{\poch{1/\alpha-1}{1/\alpha}}.
  \label{eq:mean_q_inf}
\end{equation}
It can be seen that the expectation of cluster size, under the condition 
that the in-degree is known, is finite in an unbounded network. It stands in contrast
to the unconditional cluster size, discussed in the previous section, which 
diverges logarithmically as the size of the network grows beyond every limit.

\begin{figure}[tb]
  \begin{center}
    \psfrag{q}[c][c][1]{In-degree, $q$}
    \psfrag{E(n|q)}[c][c][1]{$\meanq{\tau}{n}$}
    \psfrag{alpha=0}[r][r][1]{$\alpha=0$}
    \psfrag{alpha=1/3}[r][r][1]{$\alpha=1/3$}
    \psfrag{alpha=1/2}[r][r][1]{$\alpha=1/2$}
    \psfrag{alpha=2/3}[r][r][1]{$\alpha=2/3$}
    \psfrag{ 0}[c][c][1]{$0$}
    \psfrag{ 1}[c][c][1]{$1$}
    \psfrag{ 10}[c][c][1]{$10$}
    \psfrag{ 100}[c][c][1]{$10^2$}
    \psfrag{ 1000}[c][c][1]{$10^3$}
    \psfrag{ 10000}[c][c][1]{$10^4$}
    \psfrag{ 100000}[c][c][1]{$10^5$}
    \psfrag{ 1e+06}[c][c][1]{$10^6$}
    \resizebox{0.5\textwidth}{!}{\includegraphics{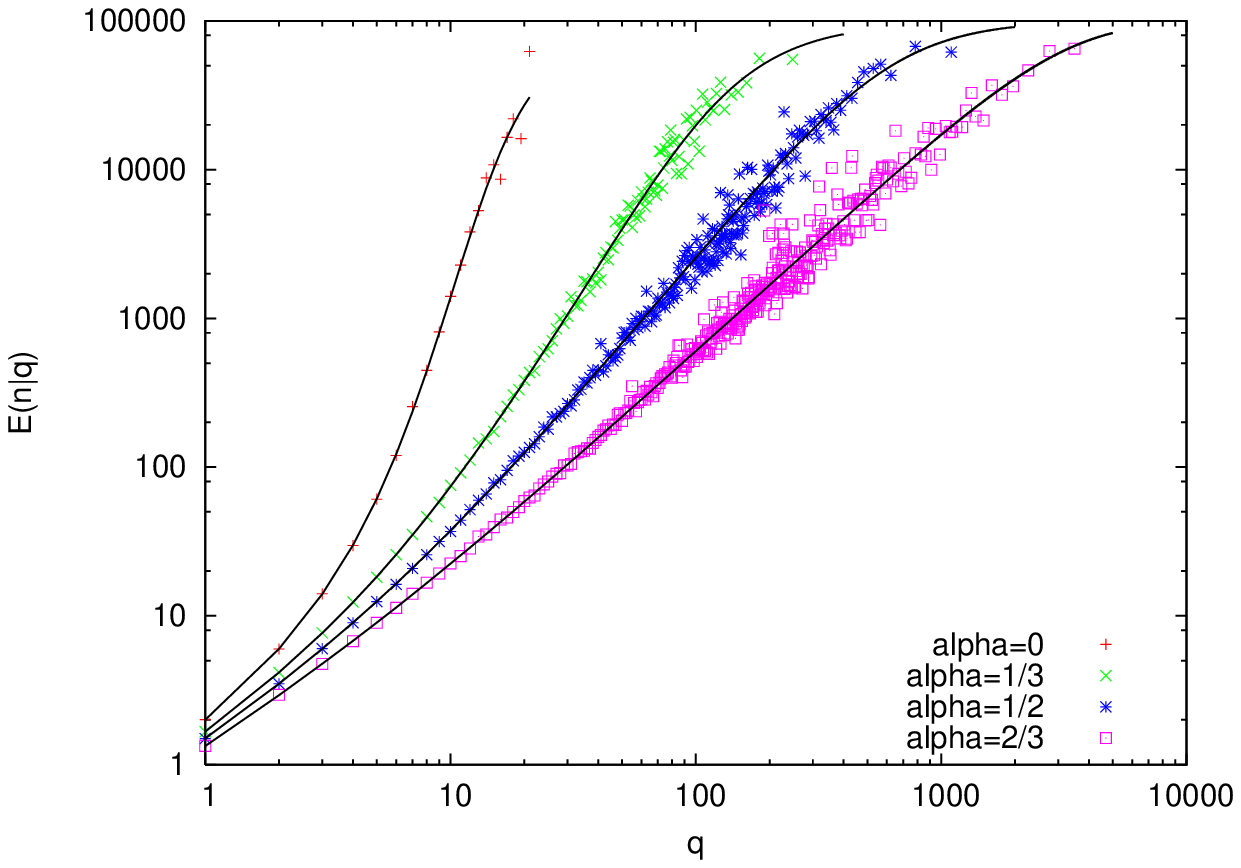}}
  \end{center}
  \caption{(Color online) Figure shows the average cluster size as the function of the in-degree 
  $q$, obtained from 100 realizations of $10^5$ size networks.
  Simulation data has been collected at $\alpha=0$, $1/3$, $1/2$, and $2/3$ parameter
  values. Analytical result (\ref{eq:mean_q}) of conditional expectation $\meanq{\tau}{n}$
  is shown with continuous lines.}
  \label{fig:deg-cluster-avg}
\end{figure}

In the ER limit, the expected conditional cluster size becomes
\begin{equation}
  \lim_{\alpha\to0}\meanq{\infty}{n+2}=2^{q+1}.
  \label{eq:mean_q_inf_a0}
\end{equation}
The fundamental difference between the scale-free and non-scale-free networks
can be observed again. In the scale-free case the expected conditional cluster size
asymptotically grows with the in-degree to the power of $1/\alpha$, while in the
later case it grows exponentially. On Figure~\ref{fig:deg-cluster-avg} 
the exact analytic formula (\ref{eq:mean_q}) is compared with simulation 
results at $\alpha=0$, $1/3$, $1/2$, and $2/3$. The simulations clearly justify 
our analytic solution.

\begin{figure}[tb]
  \begin{center}
    \psfrag{n}[c][c][1]{Cluster size, $n$}
    \psfrag{E(q|n)}[c][c][1]{$\meann{\tau}{q}$}
    \psfrag{alpha=0}[r][r][1]{$\alpha=0$}
    \psfrag{alpha=1/3}[r][r][1]{$\alpha=1/3$}
    \psfrag{alpha=1/2}[r][r][1]{$\alpha=1/2$}
    \psfrag{alpha=2/3}[r][r][1]{$\alpha=2/3$}
    \psfrag{ 0}[c][c][1]{$0$}
    \psfrag{ 1}[c][c][1]{$1$}
    \psfrag{ 10}[c][c][1]{$10$}
    \psfrag{ 100}[c][c][1]{$10^2$}
    \psfrag{ 1000}[c][c][1]{$10^3$}
    \psfrag{ 10000}[c][c][1]{$10^4$}
    \psfrag{ 100000}[c][c][1]{$10^5$}
    \psfrag{ 1e+06}[c][c][1]{$10^6$}
    \resizebox{0.5\textwidth}{!}{\includegraphics{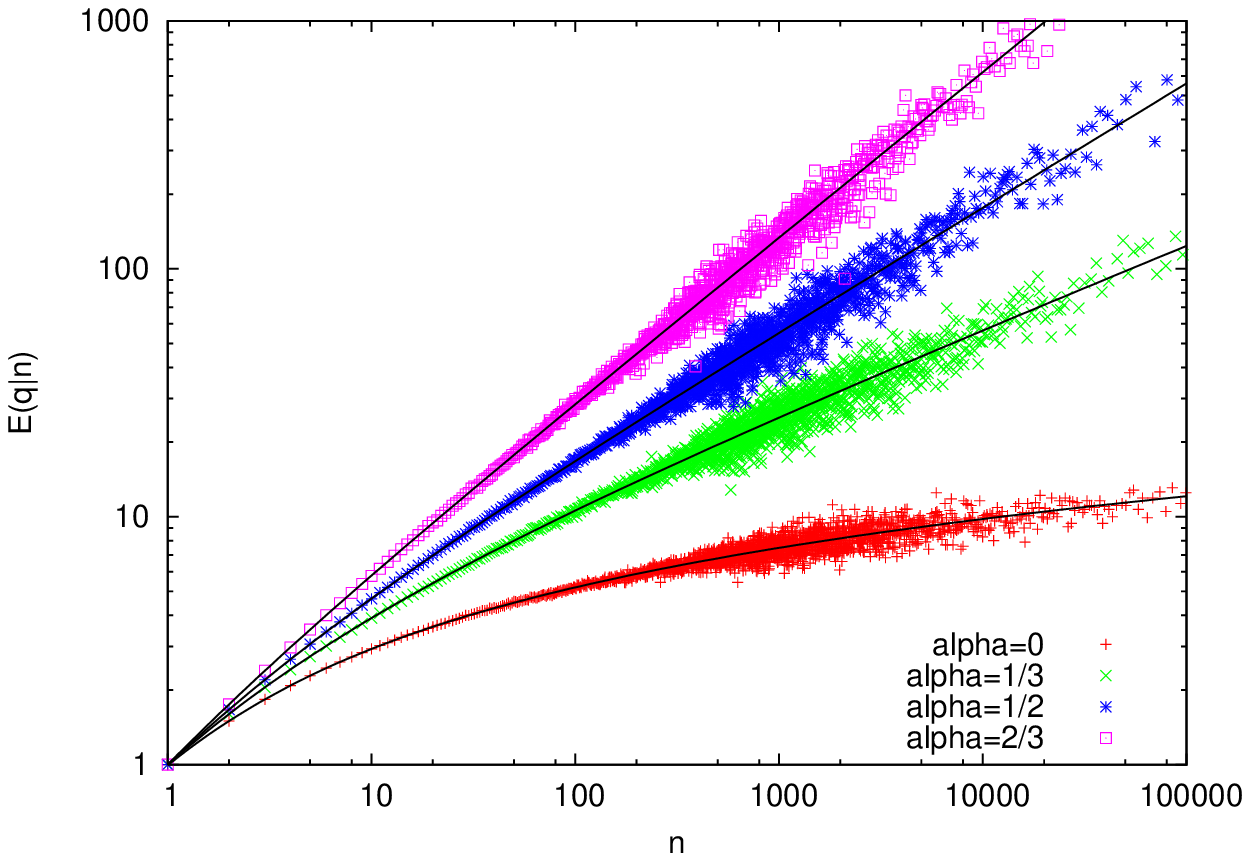}}
  \end{center}
  \caption{(Color online) Figure shows the average in-degree as the function of the cluster size 
  $n$, obtained from 100 realizations of $10^5$ size networks.
  Simulation data has been collected at $\alpha=0$, $1/3$, $1/2$, and $2/3$ parameter
  values. Analytical result (\ref{eq:mean_n}) of conditional expectation $\meann{\tau}{q}$
  is shown with continuous lines.}
  \label{fig:cluster-deg-avg}
\end{figure}

Let us investigate shortly the opposite scenario, that is when the cluster size
is known and the statistics of the in-degree under this condition is sought.
The conditional distribution can be obtained from the combination of
Eqs.~(\ref{eq:sol_joint}), (\ref{eq:sol_marginal_n}) and the definition
\begin{equation}
  \prob{\tau}{q\mid n}=\frac{\prob{\tau}{n,q}}{\prob{\tau}{n}}.
\end{equation}
The conditional expectation of in-degree can be acquired by the same technique as the
conditional expectation of cluster size. Let us calculate 
$\meann{\tau}{q+1/\alpha-1}=\meann{\tau}{q}+1/\alpha-1$ instead of $\meann{\tau}{q}$ directly:
\begin{align}
  \notag
  \meann{\tau}{q+1/\alpha-1}
  &=\frac{1}{\prob{\tau}{n}}\sum_{q=0}^{n}\left(q+1/\alpha-1\right)\prob{\tau}{n,q}\\
  &=\frac{\Gamma(2-\alpha)}{\alpha}\poch{n+1-\alpha}{\alpha},
  \label{eq:mean_n}
\end{align}
where $0\le n<\tau$. Note, that the conditional expectation of in-degree is
independent of $\tau$, that is of the size of the network. 
In the ER limit the expectation of the in-degree becomes
\begin{equation}
  \lim_{\alpha\to0}\meann{\tau}{q}=\Psi(n+1)+\gamma,
\end{equation}
where $\Psi(x)=\frac{d}{dx}\ln\Gamma(x)$ denotes the digamma function, and 
$\gamma=-\Psi(1)\approx 0.5772$ is the Euler--Mascheroni constant.
Asymptotically the expectation of the in-degree in a scale-free tree grows
with the cluster size to the power of $\alpha$, while in a ER tree it grows only
logarithmically, since $\Psi(n+1)=\log n+\Ordo{1/n}$. 
Therefore, conditional in-degree and conditional cluster
size are mutually inverses \emph{asymptotically}. Figure~\ref{fig:cluster-deg-avg}
shows the analytic solution (\ref{eq:mean_n}) and simulation data at $\alpha=0$, 
$1/3$, $1/2$, and $2/3$ parameter values. Simulation data has been collected from
100 realizations of $10^5$ size networks.

\section{Conditional distribution of edge betweenness}
\label{sec:load_prob}

Using the results of the previous sections, we are finally ready to answer 
the problem which motivated our work, that is the distribution of the
edge betweenness under the condition that the in-degree of the ``younger''
node of the link is known. It has been noted at the beginning of 
Section~\ref{sec:cond_joint_prob} that the edge betweenness can be expressed
with cluster size: 
\begin{equation}
  L=\left(n+1\right)\left(\tau-n\right).
  \label{eq:load_cluster}
\end{equation}
Therefore, conditional edge betweenness can be given formally by the following transformation
of random variable $n$:
\begin{equation}
  \prob{\tau}{L\mid q}=\sum_{n=0}^{\tau-1}\delta_{L,\left(n+1\right)\left(\tau-n\right)}\prob{\tau}{n\mid q}.
\end{equation}
Obviously, $\prob{\tau}{L\mid q}$ is non-zero only at those values of $L$, where
\eqref{eq:load_cluster} has integer solution for $n$. If 
\begin{equation}
  n_L=\frac{\tau-1}{2}-\sqrt{\frac{\left(\tau+1\right)^2}{4}-L}
  \label{eq:cluster_load}
\end{equation}
is such an integer solution of the quadratic equation \eqref{eq:load_cluster}, 
and $L\neq \left(\tau+1\right)^2/4$, then 
\begin{equation}
  \prob{\tau}{L\mid q}=\prob{\tau}{n_L\mid q}+\prob{\tau}{\tau-1-n_L\mid q}.
\end{equation} 
If $L=\left(\tau+1\right)^2/4$ is integer, then $\prob{\tau}{L\mid q}=\prob{\tau}{n_L\mid q}$.

The conditional expectation of edge betweenness can be obtained from \eqref{eq:load_cluster}:
\begin{equation}
  \label{eq:meanq_load_expand}
  \meanq{\tau}{L}=\tau\meanq{\tau}{n+1}-\meanq{\tau}{\left(n+1\right)n}.
\end{equation}
Therefore, for the exact calculation of $\meanq{\tau}{L}$ the first and the
second moment of the conditional cluster size distribution are required. The
first moment, that is the mean, has been derived in the previous section. 
In order to calculate the second moment let us use the technique we have 
developed in the previous sections. Let us consider:
\begin{equation}
  \meanq{\tau}{\left(n+2-\alpha\right)\left(n+1-\alpha\right)}
  =\frac{\tau+1-\alpha}{\tau}\frac{\poch{1/\alpha-1}{q}}{\prob{\tau}{q}}
  \sum_{n=q}^{\tau-1}\frac{\sumterm{n,q}}{\poch{2-\alpha}{n-1}}.
\end{equation}
We shall be cautious when the summation for $n$ is evaluated.
The $k=1$ term in 
$\sumterm{n,q}=\sum_{k=0}^{q}\frac{\left(-1\right)^k}{k!\left(q-k\right)!}\poch{-\alpha k}{n}$
must be treated separately to avoid a divergent term:
  \begin{align*}
    \sum_{n=q}^{\tau-1}\frac{\sumterm{n,q}}{\poch{2-\alpha}{n-1}}
    &=\frac{1-\alpha}{\left(q-1\right)!}\left[\alpha\Psi(\tau-\alpha)-\alpha\Psi(1-\alpha)-\Psi(q)-\gamma\right]\\
    &-\frac{1}{\alpha}\frac{1}{\poch{2-\alpha}{\tau-2}}\sum_{k=2}^{q}\frac{\left(-1\right)^k}{k!\left(q-k\right)!}
    \frac{\poch{-\alpha k}{\tau}}{k-1}
\end{align*}
The exact formula for $\meanq{\tau}{L}$ can be obtained straightforwardly, after
\eqref{eq:mean_q} and the above expressions have been substituted into
\eqref{eq:meanq_load_expand}. 

Let us consider the scenario when the size of the network tends to infinity.
Equation~\eqref{eq:load_cluster} implies that edge betweenness diverges as
$\tau\to\infty$, therefore $L$ should be rescaled for an infinite network.
From the asymptotics of the digamma function
$\Psi(\tau-\alpha)=\ln\tau+\Ordo{1/\tau}$ it follows that
$\meanq{\tau}{\left(n+2-\alpha\right)\left(n+1-\alpha\right)}$ grows only
logarithmically, slower than the linear growth of
$\tau\meanq{\tau}{n+2-\alpha}$. Therefore, edge betweenness asymptotically
grows linearly as the size of the network grows beyond every limit.
Let us rescale edge betweenness
\begin{equation}
  \Lambda_{\tau}=\frac{L(\tau)}{\tau+1}
\end{equation}
and let us consider the limit $\Lambda=\lim_{\tau\to\infty}\Lambda_{\tau}=n_{\Lambda}+1$.
The CCDF of the rescaled edge betweenness
can be given by $\ccdf{\infty}{\Lambda\mid q}
  =\lim_{\tau\to\infty}\sum_{n=n_{\Lambda_{\tau}}}^{\tau-1-n_{\Lambda_{\tau}}}\prob{\tau}{n\mid q}
  =\frac{1}{\prob{\infty}{q}}\sum_{n=\Lambda-1}^{\infty}\prob{\infty}{n,q}$. 
When the summation has been carried out, the following equation is obtained:
\begin{equation}
  \ccdf{\infty}{\Lambda\mid q}
  =\frac{\poch{2/\alpha-1}{q+1}}{\poch{2-\alpha}{\Lambda-1}}
  \sum_{k=0}^{q}\frac{\left(-1\right)^k}{k!\left(q-k\right)!}
  \frac{\poch{-\alpha k}{\Lambda-1}}{k+2/\alpha-1},
  \label{eq:ccdf_load}
\end{equation}
where $q+1\le\Lambda$. 
{ If $1<q\ll\Lambda$, then only the first term of the sum should be taken into
account, and it is easy to see that
\begin{equation}
  \ccdf{\infty}{\Lambda\mid q}=\frac{\alpha^2\left(1-\alpha\right)}{2\Gamma(2/\alpha-1)}
  \frac{q^{2/\alpha}}{\Lambda^2}+\Ordo{1/\Lambda^{2+\alpha}}.
  \label{eq:ccdf_load_approx}
\end{equation}
It can be seen that the scaling exponent $-2$ is independent of $\alpha$.
The above asymptotic formula has been obtained for infinite networks.
The same power law scaling can be observed in finite size networks as \eqref{eq:ccdf_load_approx}
if $\Lambda_{\tau}\ll\tau$. However, $\ccdf{\tau}{\Lambda_{\tau}\mid q}\equiv0$ 
if $\Lambda_{\tau}>\tau$ in finite networks, therefore asymptotic formula 
\eqref{eq:ccdf_load_approx} evidently  becomes invalid if 
$\Lambda_{\tau}\approx\tau$.

It is obvious that as the size of the network grows larger and larger,
asymptotic formula \eqref{eq:ccdf_load} becomes more and more
accurate. One can ask how fast the convergence is. From elementary estimations
of $\ccdf{\tau}{\Lambda_{\tau}\mid q}$ one can show that for fixed 
$\Lambda_{\tau}$:
\begin{equation}
  \ccdf{\tau}{\Lambda_{\tau}\mid q}=\ccdf{\infty}{\Lambda_{\tau}\mid q}
  -\left(1-\ccdf{\infty}{\Lambda_{\tau}\mid q}\right) 
  \frac{\alpha^2\left(1-\alpha\right)}{2}\frac{1}{\tau^2}
  +\Ordo{1/\tau^{2+\alpha}},
   \label{eq:ccdf_size_approx}
\end{equation}
that is corrections to the asymptotic formula decrease with $\tau^{-2}$
for large $\tau$.
}

\begin{figure}[tb]
  \begin{center}
    \psfrag{Lambda}[c][c][1]{Rescaled edge betweenness, $\Lambda$}
    \psfrag{Fc(L|q)}[c][c][1]{$\ccdf{\infty}{\Lambda\mid q}$}
    \psfrag{N=1e4}[r][r][1]{$N=10^4$}
    \psfrag{N=1e5}[r][r][1]{$N=10^5$}
    \psfrag{N=1e6}[r][r][1]{$N=10^6$}
    \psfrag{q=1}[r][r][1]{$q=1$}
    \psfrag{q=2}[r][r][1]{$q=2$}
    \psfrag{ 1}[r][r][1]{$1$}
    \psfrag{ 0.1}[r][r][1]{$10^{-1}$}
    \psfrag{ 0.01}[r][r][1]{$10^{-2}$}
    \psfrag{ 0.001}[r][r][1]{$10^{-3}$}
    \psfrag{ 1e-04}[r][r][1]{$10^{-4}$}
    \psfrag{ 1e-05}[r][r][1]{$10^{-5}$}
    \psfrag{ 1e-06}[r][r][1]{$10^{-6}$}
    \psfrag{ 10}[r][r][1]{$10$}
    \psfrag{ 100}[r][r][1]{$10^2$}
    \psfrag{ 1000}[r][r][1]{$10^3$}
    \psfrag{ 10000}[r][r][1]{$10^4$}
    \psfrag{ 100000}[r][r][1]{$10^5$}
    \psfrag{ 1e+06}[r][r][1]{$10^6$}
    \resizebox{0.5\textwidth}{!}{\includegraphics{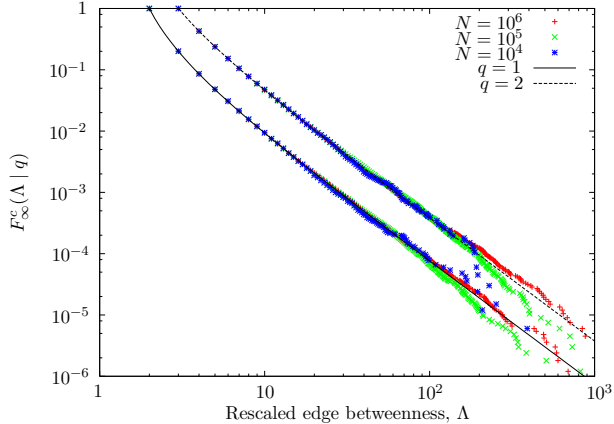}}
  \end{center}
  \caption{(Color online) Figure shows CCDF of edge betweenness under the condition that
  the in-degree $q$ is known. Empirical CCDF has been obtained from $100$ 
  realizations of $N=10^4$ and $N=10^5$, and $10$ realizations of $N=10^6$ size
  networks at $\alpha=1/2$ parameter value. Continuous lines show analytic result
  (\ref{eq:ccdf_load}).}
  \label{fig:load-dist-1}
\end{figure}

\begin{figure}[tb]
  \begin{center}
    \psfrag{q}[c][c][1]{In-degree, $q$}
    \psfrag{E(L|q)}[c][c][1]{$\meanq{\infty}{\Lambda}$}
    \psfrag{N=1e4}[r][r][1]{$N=10^4$}
    \psfrag{N=1e5}[r][r][1]{$N=10^5$}
    \psfrag{N=1e6}[r][r][1]{$N=10^6$}
    \psfrag{alpha=0}[r][r][1]{$\alpha=0$}
    \psfrag{alpha=1/2}[r][r][1]{$\alpha=1/2$}
    \psfrag{ 0}[r][r][1]{$0$}
    \psfrag{ 1}[r][r][1]{$1$}
    \psfrag{ 5}[r][r][1]{$5$}
    \psfrag{ 10}[r][r][1]{$10$}
    \psfrag{ 15}[r][r][1]{$15$}
    \psfrag{ 20}[r][r][1]{$20$}
    \psfrag{ 100}[r][r][1]{$10^2$}
    \psfrag{ 1000}[r][r][1]{$10^3$}
    \psfrag{ 10000}[r][r][1]{$10^4$}
    \psfrag{ 100000}[r][r][1]{$10^5$}
    \psfrag{ 1e+06}[r][r][1]{$10^6$}
    \resizebox{0.5\textwidth}{!}{\includegraphics{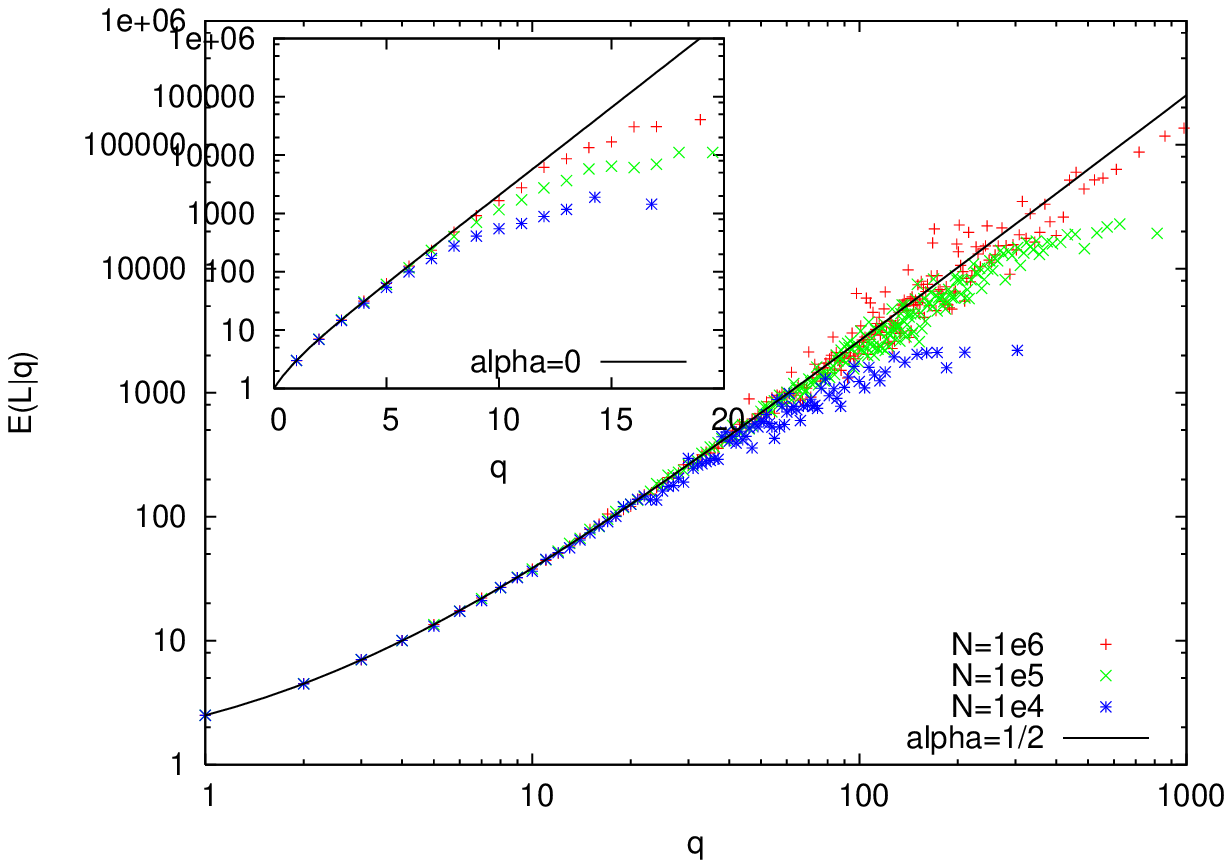}}
  \end{center}
  \caption{(Color online) Figure shows average edge betweenness under the condition 
  that the in-degree $q$ is known as the function of $q$ on log-log plot. 
  Numerical data has been collected from $100$ 
  realizations of $N=10^4$ and $N=10^5$, and $10$ realizations of $N=10^6$ size
  networks at $\alpha=1/2$ parameter value. Inset shows the same scenario at 
  $\alpha=0$ parameter value on semi-logarithmic plot. Continuous lines show analytic 
  results \eqref{eq:deg-load} and \eqref{eq:deg-load_a0}.}
  \label{fig:deg-load-avg}
\end{figure}

On Figure~\ref{fig:load-dist-1} comparison of analytic formula
\eqref{eq:ccdf_load} with simulation results is presented for $q=1$ and $q=2$.
The empirical CCDF of rescaled edge betweenness, under the condition that
in-degree $q$ is known, is shown for $10^4$, $10^5$, and $10^6$ size networks,
at $\alpha=1/2$ parameter value. The empirical CCDFs of rescaled edge
betweenness evidently collapse to the same curve for different size networks,
and they coincide precisely with our analytic result.

The expectation of the rescaled edge betweenness under the condition that in-degree
$q$ is known can be given by $\meanq{\infty}{\Lambda}=\meanq{\infty}{n_{\Lambda}+1}$.
Using \eqref{eq:mean_q_inf} and \eqref{eq:mean_q_inf_a0} we get
\begin{align}
  \label{eq:deg-load}
  \meanq{\infty}{\Lambda}
  &=\left(1-\alpha\right)\frac{\poch{q+1/\alpha}{1/\alpha}}{\poch{1/\alpha-1}{1/\alpha}}-1+\alpha,\\
  \lim_{\alpha\to0}\meanq{\infty}{\Lambda}&=2^{q+1}-1.
  \label{eq:deg-load_a0}
\end{align}
One can see that $\meanq{\infty}{\Lambda}\sim q^{1/\alpha}$ for $q\gg1$ if $\alpha>0$ and 
$\meanq{\infty}{\Lambda}\sim e^{q}$ for $q\gg1$ if $\alpha\to0$.

Analytic results \eqref{eq:deg-load} and \eqref{eq:deg-load_a0}, and simulation
data are shown in Figure~\ref{fig:deg-load-avg} at $\alpha=1/2$ and $\alpha=0$
parameter values. Numerical data has been collected from the same $10^4$,
$10^5$, and $10^6$ size networks as above. As the size of the network grows,
a larger and larger range of the rescaled empirical data collapses to the same
analytic curve. On the high degree region some discrepancy can be observed due
to the finite scale effects.

Finally, let us note that the precise unconditional distribution of edge betweenness
$\prob{\tau}{L}=\sum_{n=0}^{\tau-1}\delta_{L,\left(n+1\right)\left(\tau-n\right)}\prob{\tau}{n}$
can be obtained from \eqref{eq:sol_marginal_n} as well. Furthermore, CCDF
of the unconditional betweenness
$\ccdf{\tau}{L}=\sum_{n=n_L}^{\tau-n_L-1}\prob{\tau}{n}$ can be derived in
closed form:
\begin{equation}
  F_{\tau}^{c}(L)=\frac{\tau+1-\alpha}{\tau}
  \frac{\left(1-\alpha\right)\left(\tau-2n_L\right)}{\left(n_L+1-\alpha\right)\left(\tau-n_L+1-\alpha\right)}.
\end{equation}

For the sake of simplicity we have assumed during our calculations that
in-degrees of the ``younger'' nodes are provided. However, it is possible that
even though both two in-degrees of every link are known, we cannot distinguish
them from each other, that is we cannot tell which is the ``younger'' node. How
could we extend our results to this scenario?  Let us consider a new edge when
it is connected to the network. The in-degree of the new node is obviously $0$.
The in-degree of the other node, which the new node is connected to, is equal
to or larger than one.  Due to preferential attachment the larger the in-degree
is the faster it grows. Even if preferential attachment is absent, the growth
rate of every in-degree is the same. Therefore, it is expected that the initial
deficit in the in-degree of the ``younger'' node grows or remains at the same
level during the evolution of the network. It follows that it is a reasonable
approximation to substitute the in-degree of the ``younger'' node $q$ with
$q_{\text{min}}=\min(q_1,q_2)$ in our formulae.

\section{Conclusions}
\label{sec:conclusions}

{ A typical network construction problem is to design network
infrastructure without wasting precious resources at places where not needed.
An appropriate design strategy is if network resources are allocated
proportionally to the expected traffic. In a mean field approximation
the expected traffic is proportional to the number of shortest paths
going through a certain network element, that is the betweenness.  

The precise calculation of all the betweenness require complete information on
the network structure.  In real life, however, the number of shortest paths is
often impossible to tell because the structure of the network is not fully
known. One of the practical results of this paper is that the expectation of
edge betweenness can be estimated precisely when a limited local information on network
structure---the in-degree of the ``younger'' node---is available.

Another difficulty of network design is that the size of real networks is
finite. Moreover, the size of real networks is often so small that asymptotic
formulas can be applied only with unacceptable error. The other important
novelty of our results is that the derived formulas are exact even for finite
networks, which allows better design of realistic finite size networks.}

Various statistical properties of evolving random trees have been investigated
in this paper. We have focused on the cluster size, the in-degree and the edge
betweenness. We have considered the $m=1$ case of the BA model extended with
initial attractiveness for modeling random trees. Initial attractiveness allows
fine tuning of the scaling parameter. Moreover, in the limit of the tuning
parameter $\alpha\to0$ the applied model tends to a non-scale-free
structure, which is in many aspects similar to the classical ER model.
Therefore, we were able to investigate both the scale-free and the
non-scale-free scenario within the same framework.

First, the evolution of cluster size and in-degree of a specific edge have been
modeled as a bivariate Markov process. The master equation, associated with the
Markov process, has led us to a linear partial difference equation. An exact
analytic solution of the master equation, which satisfies the initial
conditions as well, has been found. The solution provides the joint probability
distribution of cluster size and in-degree for a specific edge.

Using the above results we have derived the joint probability distribution of
cluster size and in-degree for a randomly selected edge. It is of more practical
importance than the joint distribution for a specific edge because, in
contrast to the former distribution, it provides the statistical description of
the whole network. We also derived the joint distribution in the ER limit. Note
that the obtained formulae are exact for even finite size networks. In
addition, the formulae for unbounded networks have been presented as well.

We have continued our analysis with the one dimensional marginal distributions.
We have shown some fundamental differences in the scaling properties of the 
marginal cluster size and in-degree distributions.  The novelty of our results
here, compared to previous results in the literature, is that we have found
exact analytic formulae not only for the large, but also for the small cluster
size and in-degree region.

Although the marginal distributions have their own importance, we have derived
them in order to obtain conditional probability distributions. From the
combination of the joint and the marginal distributions we have given the
conditional distributions of cluster size and in-degree. We have also
presented conditional expectations of cluster size and in-degree for both
finite and unbounded networks.  We have found that asymptotically the
conditional cluster size grows with in-degree to the power of $1/\alpha$ and
the conditional in-degree grows with cluster size to the power of $\alpha$,
respectively. The ER limit has been discussed as well. We have shown that the
conditional cluster size grows exponentially and the conditional in-degree grows
logarithmically when $\alpha\to0$.

Finally, by applying the transformation of random variables we have derived the
distribution of edge betweenness under the condition that the corresponding
in-degree is known. We have found that the conditional expectation of edge
betweenness grows linearly with the size of the network. For the analysis of
unbounded networks we have defined the rescaled edge betweenness $\Lambda$, and
derived its distribution and expectation under the condition that in-degree $q$
is provided. Our analytic results have been verified at different network sizes
and parameter values by extensive numerical simulations. We have demonstrated
that numerical simulations fully confirm our analytic results.

For the future, we hope that the methods we have developed in this paper allow
us to describe cluster size and edge betweenness in more general scenarios. For
example, when not only the ``younger'', but both two in-degrees of links are
considered. 

\section*{Acknowledgement}

{ The authors thank the partial support of the National Science
Foundation (OTKA T37903), the National Office for Research and Technology (NKFP
02/032/2004 and NAP 2005/ KCKHA005) and the EU IST FET Complexity EVERGROW
Integrated Project.} The authors also thank M\'at\'e Mar\'odi for the fruitful
discussions and his helpful comments. 

\appendix

\section{Expansion of the Kronecker-delta function}
\label{app:kronecker_exp}
We have seen that the general solution of Eq.~(\ref{eq:master_eq}) is
$\probi{\tau}{n,q}=\sum_{\lambda_1,\lambda_2}C_{\lambda_1,\lambda_2}f(\tau)g(n)h(q)$,
and the initial condition is $\probi{\tau_e}{n,q}=\delta_{n,0}\delta_{q,0}$, where
\begin{equation}
  \delta_{n,m}=
  \begin{cases}
    1, &\textrm{if $n=m$},\\
    0, &\textrm{if $n\neq m$}
  \end{cases}
\end{equation}
is the Kronecker-delta function, and $n$ and $m$ are integers.
Coefficients $C_{\lambda_1,\lambda_2}$ are calculated in this 
section.  First we show that
\begin{equation}
  \label{eq:delta_exp}
  \delta_{n,0}=\sum_{k=0}^n\frac{\left(-1\right)^k}{k!}\frac1{\Gamma(n-k+1)}.
\end{equation}
Note that we can consider $m=0$ without any loss of generality, since
$\delta_{n,m}\equiv\delta_{n-m,0}$. 

If $n<0$, then the summand in (\ref{eq:delta_exp}) is zero by
definition, indeed. If $n>0$, then
\begin{equation}
  \sum_{k=0}^n\frac{\left(-1\right)^k}{k!}\frac1{\Gamma(n-k+1)}
  =\frac1{n!}\sum_{k=0}^n\binom{n}{k}\left(-1\right)^k=0
\end{equation}
follows from the binomial theorem. Finally, for $n=0$,
\begin{equation}
  \sum_{k=0}^0\frac{\left(-1\right)^k}{k!}\frac1{\Gamma(-k+1)}=
  \frac{\left(-1\right)^0}{0!}\frac1{\Gamma(1)}=1.
\end{equation}

Coefficients $C_{\lambda_1,\lambda_2}$ can be obtained from the
term by term comparison of $\probi{\tau_e}{n,q}
=\sum_{\lambda_1,\lambda_2}C_{\lambda_1,\lambda_2}f(\tau_e)\,g(n)\,h(q)$
with the expansion of the initial condition $\delta_{n,0}\,\delta_{q,0}$,
shown above. One can easily confirm with the help of identity 
$f(n)\delta_{n,0}\equiv f(0)\delta_{n,0}$ that the same terms appear on 
both sides, if $\lambda_1=-k_1$, and 
$\lambda_2=-\alpha k_2$, and coefficients $C_{k_1,k_2}$ are the following:
\begin{equation}
  C_{k_1,k_2}=\frac{\left(-1\right)^{k_1+k_2}}{k_1!\,k_2!}
  \frac{\Gamma(\tau_e+1-\alpha)}{\Gamma(\tau_e-k_1)}\frac1{\Gamma(-\alpha k_2)}
  \frac{1}{\Gamma(1/\alpha-1)}.
\end{equation}

Finally, to obtain (\ref{eq:sol_joint}) the summation for $k_1$ can be 
carried out explicitly:
\begin{align*}
  \sum_{k_1=0}^{n}\frac{\left(-1\right)^{k_1}}{k_1!\Gamma(n-k_1+1)}\frac{\Gamma(\tau-k_1)}{\Gamma(\tau_e-k_1)}
  =\frac{\Gamma(\tau-\tau_e+1)}{\Gamma(n+1)\Gamma(\tau_e)}\frac{\Gamma(\tau-n)}{\Gamma(\tau-\tau_e-n+1)}
\end{align*}

\section{The $\alpha\to0$ limit of joint distribution $\prob{\tau}{n,q}$}
\label{app:ER_limit}

In this section we prove that the ER limit of the joint probability $\prob{\tau}{n,q}$
is (\ref{eq:sol_joint_a0}).

\textbf{Theorem}
\textit{
  Let us consider $\prob{\tau}{n,q}$ as defined in (\ref{eq:sol_joint}), where
  $0<q<n<\tau$ are integers. Then the following limit holds:
  \begin{equation}
    \lim_{\alpha\to0}\prob{\tau}{n,q}=\frac{\tau+1}{\tau\Gamma(n+3)}
    \sum_{k=q-1}^{n-1}\left(-1\right)^{n-1-k}
    S_{n-1}^{\left(k\right)}\binom{k}{q-1}.
  \end{equation}
  \label{thm:ER_limit}
  }
  
\begin{proof}
First, let us note that $\sumterm{n,q}$ in (\ref{eq:sol_joint}) can be
rewritten in the following equivalent form: $\sumterm{n,q}=\alpha\sum_{k=0}^{q-1}
\frac{\left(-1\right)^k\poch{1-\alpha-\alpha k}{n-1}}{k!\left(q-1-k\right)!}$.
Next, Pochhammer's symbol $\poch{1/\alpha-1}{q}$ is replaced with its asymptotic form:
$\poch{1/\alpha-1}{q}=1/\alpha^q\left(1+\Ordo{\alpha}\right)$.
After the obvious limits have been evaluated the following equation is obtained:
\begin{equation}
  \lim_{\alpha\to0}\prob{\tau}{n,q}=\frac{\tau+1}{\tau\Gamma(n+3)}
  \lim_{\alpha\to0}\frac{
    \sum_{k=0}^{q-1}\frac{\left(-1\right)^k\poch{1-\alpha-\alpha k}{n-1}}{k!\left(q-1-k\right)!}}
  {\alpha^{q-1}}.
\end{equation}

The above limit, by definition, can be substituted with $q-1$ order differential at $\alpha=0$, 
if all the lower order derivates of the sum are zero at $\alpha=0$. Indeed,
\begin{align*}
  \lim_{\alpha\to0}\frac{
    \sum_{k=0}^{q-1}\frac{\left(-1\right)^k\poch{1-\alpha-\alpha k}{n-1}}{k!\left(q-1-k\right)!}}
  {\alpha^{q-1}}
  &=\frac{1}{m!}\frac{d^m}{d\alpha^m}
  \sum_{k=0}^{q-1}\frac{\left(-1\right)^k\poch{1-\alpha-\alpha k}{n-1}}{k!\left(q-1-k\right)!}
  \Biggr|_{\alpha=0}\\
  &=\frac{1}{m!}\frac{d^m\poch{1+\alpha}{n-1}}{d\alpha^m}
  \Biggr|_{\alpha=0}
  \sum_{k=0}^{q-1}\frac{\left(-1\right)^k\left(-k-1\right)^m}{k!\left(q-1-k\right)!},
\end{align*}
where the sum is $0$ if $m<q-1$ and $1$ if $m=q-1$. Therefore, the limit can be
transformed to
\begin{equation}
  \lim_{\alpha\to0}\prob{\tau}{n,q}=\frac{\tau+1}{\tau\Gamma(n+3)}\frac{1}{\left(q-1\right)!}
  \frac{d^{q-1}\poch{1+\alpha}{n-1}}{d\alpha^{q-1}}\biggr|_{\alpha=0}.
\end{equation}
Finally, let us consider the power expansion of Pochhammer's symbol:
$\poch{x}{m}=\sum_{k=0}^{m}\left(-1\right)^{n-k} S_m^{\left(k\right)}x^k$,
where $S_m^{\left(k\right)}$ are the Stirling numbers of the first kind.
The expansion formula has been applied at $x=1+\alpha$ and $m=n-1$, which implies
\begin{align*}
  \lim_{\alpha\to0}\prob{\tau}{n,q}
  &=\frac{\tau+1}{\tau\Gamma(n+3)}\sum_{k=q-1}^{n-1}\frac{\left(-1\right)^{n-1-k}S_m^{\left(k\right)}}
  {\left(q-1\right)!}\frac{d^{q-1}\left(1+\alpha\right)^k}{d\alpha^{q-1}}\biggr|_{\alpha=0}\\
  &=\frac{\tau+1}{\tau\Gamma(n+3)}\sum_{k=q-1}^{n-1}\left(-1\right)^{n-1-k}S_{n-1}^{\left(k\right)}\binom{k}{q-1}.
\end{align*}
\end{proof}

\end{document}